\documentclass[aps,prl,showpacs,twocolumn]{revtex4}
\usepackage{amssymb}
\usepackage{epsfig}
\usepackage{graphicx}
\usepackage{amsmath}
\usepackage{array,color}
\usepackage{dcolumn}
\usepackage{bm}

\everymath={\fam0 }

\begin{document}

\title{Negative Magnetoresistance and Spin Filtering of Spin-Coupled Diiron-Oxo Clusters}

\author{Rui-Ning Wang$^1$}
\author{Jorge H. Rodriguez$^2$}
\author{Wu-Ming Liu$^{1}$}
\affiliation{
$^1$Beijing National Laboratory for Condensed Matter Physics, Institute of Physics, Chinese Academy of Sciences, Beijing 100190, China\\
$^2$Department of Physics, Purdue University, West Lafayette, Indiana 47907-2036, USA}

\date{\today}

\begin{abstract}

Spin dependent transport has been investigated for an {\it open shell singlet} diiron-oxo cluster. Currents and magnetoresistances have been studied, as a function of spin state, within the non-equilibrium Green's function approach. The applied bias can be used for tuning the sign of the observed magnetoresistance. A colossal magnetoresistance ratio has been determined, on the order of to 6000$\%$, for hydrogen anchoring. Applied biases lower than 0.3 V, in conjunction with sulfur anchoring, induce a negative magnetoresistance due to lowering of the anchor-scatterer tunneling barrier. In addition, the diiron-oxo cluster displays nearly perfect spin filtering for parallel alignment of the iron magnetic moments due to energetic proximity, relative to the Fermi level, of its highest occupied molecular orbitals.

\end{abstract}


\maketitle

$\it{Introduction}$.-Miniaturization of electronic devices can be aided by innovations within two fairly novel disciplines, namely spintronics \cite{science.294.1488, rmp.80.1517, prb.72.172411} and molecular electronics \cite{prb.68.035416, cpl.29.277, nature.408.541}. Initially, these two disciplines did not overlap significantly \cite{adma.23.1583, prl.96.076604, prl.102.106401}. However, more recently, a link has been established between spintronics and molecular electronics as interest has grown in the use of magnetically ordered molecular clusters for electronic transport \cite{nm.4.335, nm.10.502, nc.424.654, jacs.134.777, jmc.19.1696, apl.89.212104}. For example, nanosized Fe- or Mn-based single molecule magnets (SMMs), which can potentially serve as magnetically-functional units or logic devices, have been extensively studied \cite{nm.7.179, nature.365.141, prb.80.104422, prl.96.206801, prl.102.246801}. In particular, electron transport properties of the prototypal SMM Mn$_{12}$ \cite{ac.36.2042}, a ferrimagnetic cluster with polymetallic core surrounded by organic ligands, have been studied experimentally \cite{prl.96.206801} and theoretically \cite{prl.96.206801, prl.102.246801, jap.105.07E309}. Mn$_{12}$ has a ground state with rather large spin (S=10) \cite{jacs.113.5873, jacs.117.301} arising from ferrimagnetic ordering of its two sublattice magnetizations.

By contrast to the widely studied SMMs \cite{prb.78.155403, prb.81.125447, prl.92.137203}, a different class of metal clusters have not been extensively studied in terms of their electronic transport properties. Namely, spin polarized oxygen-bridged (i.e. $\mu$-O) bimetallic clusters whose S=0 ground state arises from antiparallel ordering of their two metal-centered magnetic moments. These systems are {\it open shell singlets} since their ground state multiplicity is M=2S+1=1 but their spin density, $\rho^{S}(\vec{r})=\rho^{\uparrow}(\vec{r})-\rho^{\downarrow}(\vec{r})$, is finite and of opposite polarity in the proximity of each of their two metal centers \cite{jcp.116.6253}. Contrary to SMMs, the spin polarized bimetallic clusters studied herein do not have an intrinsic magnetic anisotropy barrier due to their S=0 ground state. Accordingly, we study the representative {\it open shell singlet} shown in Fig. 1, herein referred as Fe$_{A}^{3+}$-($\mu$-O)-Fe$_{B}^{3+}$ or {\it diiron-oxo} cluster \cite{jacs.105.4837}, whose iron ions have nominal Fe$_{i}^{3+}$ oxidation state, S$_i=5/2$ spin (i=A, B), and antiparallel magnetic moments (AP) leading to a net $S=S_A-S_B=0$ ground state (Fig. 1) \cite{jacs.105.4837, jcp.116.6253}. Each iron ion constitutes a spin center indirectly interacting with the other {\it via} their common $\mu$-O ligand. The latter mediates superexchange interactions \cite{jcp.116.6253} which couple the irons antiferromagnetically. The lowest and highest eigenstates of the corresponding Heisenberg Hamiltonian ($\mathcal{H}_{\mathcal{HB}}=J\vec{S}_{A}\bullet{\vec{S}_B}$) are shown in Fig. 1 where the exchange constant, $J \approx +242 \hspace{2mm} cm^{- 1}$, has been determined from magnetic susceptibility \cite{jacs.106.3653}. In addition to elucidating their intricate electronic transport properties, study of such {\it diiron-oxo} clusters is of biological interest since their metallic cores constitute magneto-structural models for active sites in {\it diiron-oxo} proteins \cite{cr.90.585}.

In this Letter, we study the spin polarized transport properties of the antiparallel magnetic moment state as the cluster is connected to gold electrodes {\it via} hydrogens. The parallel moment (P) state is 15J higher in energy with a net spin S=5. Thus, it was  also of interest to study the transport properties of P states. For comparison, the cluster was also connected to gold electrodes {\it via} thiol groups, which are commonly used to
\begin{figure}[hbp]
\includegraphics[bb=30 15 190 127, width=0.35\textwidth]{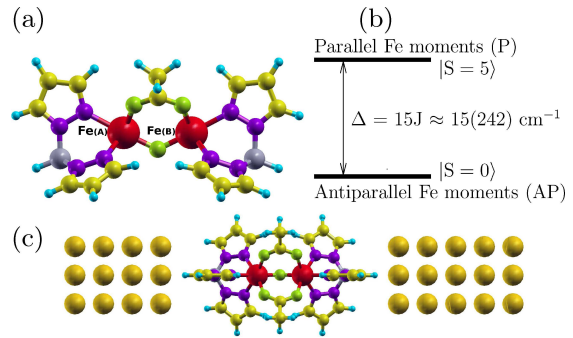}
\caption{(Color online) (a) Side view of a Fe$_A^{3+}$-($\mu$-O)($\mu$-O$_2$CCH$_3$)$_2$-Fe$_B^{3+}$(HBpz$_3$)$_2$
cluster. (b) Highest, $|S=S_A+S_B=5 \rangle$, and lowest, $|S=S_A-S_B=0 \rangle$, eigenstates of the Heisenberg Hamiltonian. (c) Top view of cluster in contact with gold electrodes. Atoms colors: iron (red), oxygen (green), nitrogen (purple), carbon (gray), hydrogen (blue).}
\label{structure}
\end{figure}
\begin{figure}[htp]
\includegraphics[bb=170 30 650 540,width=0.3\textwidth]{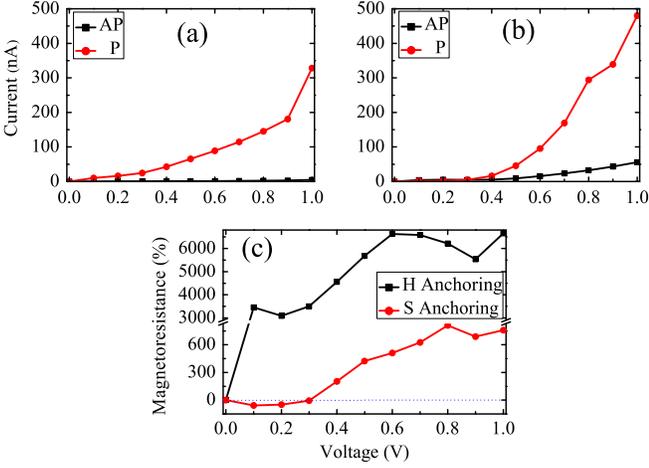}
\caption{(Color online). Transport properties of Fe$^{3+}$-($\mu$-O)-Fe$^{3+}$(HBpz$_3$)$_2$ in contact with gold electrodes. Current-voltage profile with hydrogen (a) or sulfur (b) atoms as anchoring groups. (c) Magnetoresistance for sulfur and hydrogen anchoring.}
\label{nmr}
\end{figure}
build strong chemical links \cite{science.301.1221, prb.83.064423}.
Our calculations show that the P configuration exhibits a spin-filtering effect \cite{pre.76.026605, prb.84.035323} for both types of anchors. Finally, a negative magnetoresistance has been determined below a bias of 0.3 V \cite{prl.109.206601, nl.13.481, prb.62.3010} which originates in a reduced tunneling barrier.

$\it{Methods}$.-A junction was constructed in which the central scatterer, Fe$_A^{3+}$-($\mu$-O)-Fe$_B^{3+}$, was modeled in both (AP and P) magnetic configurations (Fig. 1). The {\it diiron-oxo} cluster was placed between two atomic scale gold leads of finite cross section along the (100) direction. Terminal atoms were symmetrically anchored at the hollow site of the surface of the two leads. In an initial step, the system was partially optimized {\it via} spin polarized density functional theory (SDFT) \cite{jpcm.14.2745} by keeping the gold electrodes frozen. Only minor differences in the cluster's geometric structure were found upon optimization with hydrogen- and sulfur-anchoring. A double-zeta plus polarization basis and the generalized gradient approximation parameterized by Perdew, Burke, and Ernzerhof \cite{prl.77.3865} were used. Geometrical convergence was achieved when the forces were less than 0.03 $eV/\AA$. Nonlocal norm conserving pseudopotentials \cite{prl.43.1494} were constructed and used with scalar relativistic terms and core corrections following Troullier and Martins \cite{prb.43.1993}. In a second stage, electronic transport calculations were carried out with SMEAGOL \cite{prb.73.085414} which combines SDFT with the non-equilibrium Green's function formalism \cite{prb.50.5528, prb.63.245407, prl.82.398}.

\begin{figure}[tp]
\includegraphics[bb=200 30 640 580, width=0.25\textwidth]{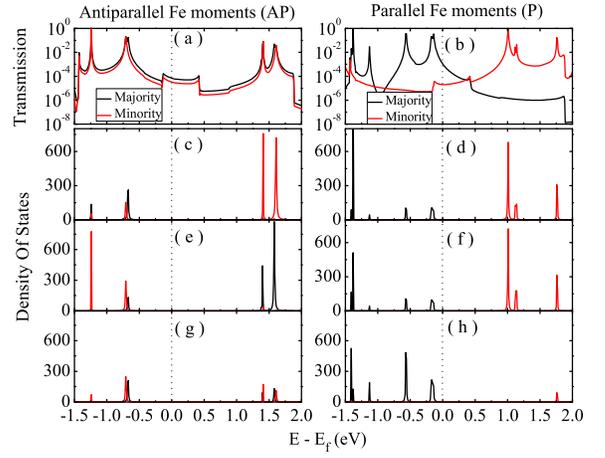}
\caption{(Color online.) Hydrogen anchoring: (a,b) Logarithm of the spin-resolved transmission spectra. Partial density of states (DOS):(c, d) Fe$_A$; (e, f) Fe$_B$; (g, h) $\mu$-O.}
\label{h-transmission}
\end{figure}

$\it{Colossal\hspace{1 mm} and \hspace{1 mm}Negative \hspace{1 mm} Magnetoresistances.}$-Spin polarized currents were self-consistently calculated within the non-equilibrium Green's function approach with the voltage-dependent Landauer-B\"{u}ttike formula,
$$I_\sigma\! =\!\frac{e}{h}\int^{\mu_{R}}_{\mu_{L}}T_{\sigma}(E,V)[f_L(E\! -\!\mu_L)\! -\!f_R(E\! -\!\mu_R)]dE,$$ where $\sigma$=$\uparrow$ or $\downarrow$, f$_{L,R}$(E)=1/(1+$e^{\frac{E-\mu_{L,R}}{k_{B}T}})$ are the Fermi-Dirac distributions, and $\mu_{L,R}$ are the chemical potentials for left and right electrodes, respectively. $T_{\sigma}(E,V)\!=\!Tr[Im(\Sigma^r_L)G^rIm(\Sigma^r_R)G^a]$ are the spin-dependent transmission coefficients. Self-consistent currents for AP and P configurations of Fe$_A^{3+}$-($\mu$-O)-Fe$_B^{3+}$ in contact with gold electrodes with hydrogen- and sulfur-anchoring are shown in Figs. 2(a,b), respectively. The currents of the higher energy P states were about two (hydrogen-anchoring) or one (sulfur-anchoring) orders of magnitude larger than those of their corresponding AP states for a 1.0 V bias. Thus, overall, the parallel moment states displayed much better conductivity. This is consistent with the high degree of Fe$_A$$\leftrightarrow$($\mu$-O)$\leftrightarrow$Fe$_B$ delocalization of the {\it diiron-oxo} cluster's highest occupied molecular orbitals (HOMOs) in its parallel configuration \cite{jcp.116.6253}. The development of spintronic devices based on giant magnetoresistances (MR) \cite{prl.61.2472, prb.81.165404, prb.82.184402} has revolutionized the magnetic memory industry. From the predicted current-voltage profiles we can infer a MR ratio defined in terms of the currents associated with AP and P states, $MR=(I_{P}-I_{AP})/I_{AP}$, which imposes no upper bound and is shown in Fig. 2(c). For hydrogen-anchoring, the MR ratio increased from 0.2 V to 0.6 V and reached a maximum ($\sim 6000\%$) at 0.6 V. Thus, the {\it diiron-oxo} cluster displayed a colossal MR effect. For sulfur-anchoring, the MR ratio was less than zero in the low bias zone (0 - 0.3 $V$) and the {\it diiron-oxo} junction showed a negative MR effect indicating that the current associated with the AP state is greater than that of the P state. Above a 0.3 V bias, the MR ratio tended to increase with increasing bias for sulfur-anchoring. However, the MR ratio for
\begin{figure}[htp]
\includegraphics[bb=200 30 640 580, width=0.25\textwidth]{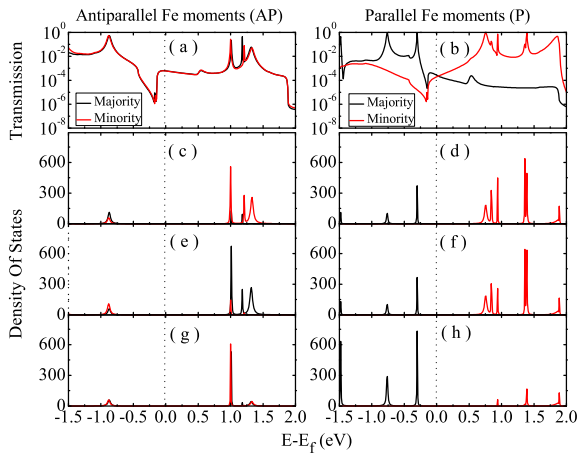}
\caption{(Color online.) Sulfur anchoring: (a,b) Logarithm of the spin-resolved transmission spectra. Partial density of states (DOS):(c, d) Fe$_A$; (e, f) Fe$_B$; (g, h) $\mu$-O.}
\label{s-transmission-rev1}
\end{figure}
sulfur-anchoring was somewhat lower than that of hydrogen-anchoring which quickly increased to a saturation value. The colossal MR effect associated with hydrogen-anchoring can be explained in terms of the transmission spectra and the cluster's electronic structure as shown in Fig. \ref{h-transmission}.

The significant difference between transport properties of AP and P states may be rationalized in terms of their transmission spectra. Figs. 3(a,b) exhibit features which strongly depend on the relative orientation (AP or P) of the magnetic moments of the irons. For both magnetic configurations the transmission probabilities decay rapidly when the energy falls below that of the lowest unoccupied molecular orbital (LUMO) or rises above that of the HOMOs. The molecular orbitals are fairly localized within the {\it diiron-oxo} cluster because it is weakly coupled to gold electrodes \cite{prb.76.224413}. In the weak-coupling limit, transport occurs when one molecular orbital is shifted to the Fermi level (E$_f$) of the leads. The molecular orbital energies are dependent on the gate voltage \cite{jacs.132.2914}. In the present system, such effect is obtained by changing the relative orientation of the magnetic moments of the iron ions. Parallel alignment, in conjunction with hydrogen-anchoring, decrease the HOMO-LUMO gap from 2.12 (for AP) to 1.15 $eV$. Reminiscent of scanning tunneling spectroscopy observations for Mn$_{12}$ SMMs \cite{prb.78.155403}, the electronic transport through the {\it diiron-oxo} cluster mainly occurs {\it via} the (P) HOMO which is in energetic proximity ($-$0.14 eV) to E$_f$ (Table I). Accordingly, the P configuration has much better conductivity properties than the AP configuration whose
\begin{table}[hlbp]
\centering \caption{The energy levels of the HOMO and LUMO of diiron-oxo clusters for AP or P alignment of magnetization with hydrogen- or sulfur-anchoring. The ratio of the majority-spin to minority-spin transmission, (T$_\uparrow$-T$_\downarrow$)/T$_\downarrow$, is computed at the energy level corresponding to HOMO.}
\begin{ruledtabular}
\begin{tabular}{*{5}{c}}
          & \multicolumn{2}{c}{H Anchoring} & \multicolumn{2}{c}{S Anchoring}\\
\hline
              &AP     &P        &AP      &P\\
\hline
 HOMO (eV)   &$-$0.71 &$-$0.14    &$-$0.87 &$-$0.30\\
 LUMO (eV)   &$+$1.41 &$+$1.01    &$+$1.01 &$+$0.76\\
 (T$_\uparrow$-T$_\downarrow$)/T$_\downarrow$ & 4.29$\times10^{-2}$ & 2.35$\times10^{4}$ & 1.65$\times10^{-1}$ & 5.50$\times10^{4}$\\
\end{tabular}
\end{ruledtabular}
\end{table}
\begin{figure}[htbp]
\includegraphics[bb=170 50 670 555, width=0.30\textwidth]{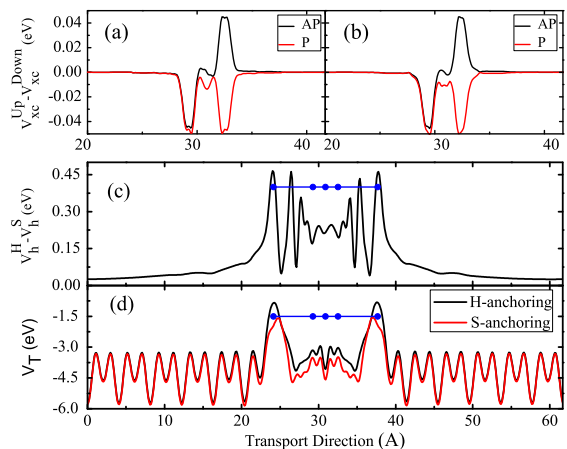}
\caption{(Color online) Planar average potentials along the transport direction of the diiron-oxo molecular junction. (a, b) Difference in exchange-correlation (xc) potentials associated with spin up and spin down electrons, V$^{Up}_{xc}$-V$^{Down}_{xc}$, for hydrogen and sulfur anchoring, respectively. (c) Difference in Hartree potentials for hydrogen and sulfur anchoring (V$^{H}_{h}$-V$^{S}_{h}$). (d) Total potentials, V$_T$, of the diiron-oxo cluster in contact with the electrodes through hydrogen and sulfur anchoring. Blue dots
indicate the locations of the left anchoring atom, left iron, the linking oxygen, the right iron, and the right anchoring atom.}
\label{potential}
\end{figure}
HOMO is at $-$0.71 eV. Therefore, Fe$_A^{3+}$-($\mu$-O)-Fe$_B^{3+}$
shows an essentially perfect colossal MR with potential use as a magnetic memory unit.

For sulfur-anchoring, the P configuration leads to features similar to those of hydrogen-anchoring. Namely, the HOMO-LUMO gap decreases from  1.88 eV in AP arrangement to 1.06 eV and the HOMO level is shifted, relative to E$_f$, from $-$0.87 to $-$0.30 eV (Fig. 4, Table I). However, for sulfur-anchoring, a negative MR effect is observed which is not directly understood from the transmission spectra alone. For SDFT calculations, the self-consistently calculated properties are dependent on the exchange-correlation (xc) and other potentials. Therefore, to rationalize the negative MR, one can consider the xc potentials associated with spin up (V$_{xc}^{Up}$) and spin down (V$_{xc}^{Down}$) electrons. The planar average potentials along the transport direction are shown in Fig. 5. Compared to AP alignments, the P configurations result in similar changes of xc potentials when {\it diiron-oxo} cluster is anchored through hydrogen- or sulfur-anchoring  as shown in Fig. 5(a,b). For AP alignment, along the transport direction, the difference xc potential (V$_{xc}^{Up}$-V$_{xc}^{Down}$), first meets a shallow potential well and then a high potential barrier. By contrast, for P alignment, the difference xc potential first meets a low potential barrier and then a deep potential well.
\begin{figure}[htp]
\includegraphics[bb=190 55 610 550, width=0.25\textwidth]{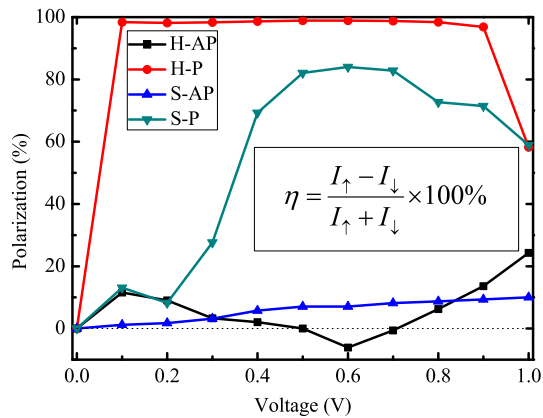}
\caption{(Color online) The polarization of the current through the
Fe$^{3+}$-($\mu$-O)-Fe$^{3+}$(HBpz$_3$)$_2$ with hydrogen and sulfur anchoring,
respectively.}
\label{magnet}
\end{figure}
Therefore, for P alignment, the majority (spin up) carriers will scatter on two potential wells around the irons during conduction. In contrast, the minority carriers traverse through two potential barriers. The absolute magnitude of the main potential wells and barriers are nearly the same for AP and P configurations due to the nominal C$_2v$ molecular symmetry.

The total potential includes, in addition to the Hartree (h) and exchange-correlation (xc) components, the pseudopotential (pseudo) representing core electrons (V$_T$=V$_h$+V$_{xc}$+V$_{pesudo}$). The negative MR effect observed for sulfur-anchoring may be attributed to the corresponding total potential becoming more negative (within the junction region) relative to hydrogen-anchoring as shown in Fig. 5(d). The current through the molecular junction not only depends on the intrinsic electronic structure of the {\it diiron-oxo} cluster but also on its, anchor dependent, interfacial barrier. When hydrogen-anchoring is replaced with sulfur-anchoring, the current increases (Fig. 2) due to the lower total interfacial potential. However, for AP alignment, such anchor-dependent potential decrease makes the current increase by about one order of magnitude in the higher bias region ($\leq$ 1.0 V) up to $\approx$50 nA. For P alignment, in the higher bias region ($\leq$ 1.0 V), the same anchoring substitution gives rise to a lesser relative increment in the current (about a factor of two) but to a much greater absolute increment, roughly from 300 to 500 nA (Fig. 2). Relative to the low bias region ($\geq$ 0.0 V), the net increment in current as a function of bias is much more pronounced for AP$\rightarrow$P realignment than for H$\rightarrow$S substitution. An AP$\rightarrow$P reorientation of the Fe moments is concomitant with substantial changes in the cluster's electronic structure \cite{jcp.116.6253}. Therefore, the current is dominated by the anchor-dependent tunneling barrier at lower voltages and by the magnetic alignment (AP or P) at higher voltages.

$\it{Spin \hspace{1mm} Valve \hspace{1mm} Effect.}$-Figures 3-4 show that, near E$_f$, the transmission for majority- and minority-spin are essentially equal for AP configurations. This is true for either type of anchoring. For P states, however, the transmissions near E$_f$ for majority-spin dominate. This effect is further illustrated by the ratios of majority- to minority-spin transmission coefficients, (T$_\uparrow$-T$_\downarrow$)/T$_\downarrow$, computed at the energy of the HOMO level. Table I shows that these ratios for P configurations are orders of magnitude larger than those of their AP counterparts. This is consistent with the partial DOS plots displaying intensities closer to E$_f$ for P states as compared to AP states. The previous features of the transmission spectra point to preferential transmission of majority-spin electrons and thus to effective spin filtering. The polarization of the current can also be analyzed in terms of the spin-injection factor $\eta=(I_\uparrow-I\downarrow)/(I_\uparrow+I_\downarrow)\times100\%$.
The polarization of the current as a function of bias and anchor are shown in Fig. \ref{magnet}. For AP states the currents are not strongly spin polarized. However, $\eta$ significantly increases for P states, particularly for hydrogen-anchoring where $\eta$ is on the order of $\sim 95\%$ in the bias range 0.1 - 0.9 V. This indicates that Fe$_A^{3+}$-($\mu$-O)-Fe$_B^{3+}$ in its P configuration may be used in the context of molecular spintronics to achieve an excellent spin filtering effect. This is particularly true at higher temperatures for which the P configurations is significantly populated.

$\it{Conclusion}$.-Either colossal or negative magnetoresistances have been computed for the sulfur-anchored {\it diiron-oxo} cluster depending on the applied bias. A spin-valve effect has been predicted for the parallel moment states, in particular for hydrogen-anchoring for which the polarization of the current is on the order of 95$\%$ over a range of applied bias. This suggests achieving control of different transport properties of the same molecular cluster by tuning the applied bias in the context of molecular spintronic devices.

$\it{Acknowledgments.}$-This work was partially supported by NKBRSFC (WML) under grants Nos. 2011CB921502, 2012CB821305, 2009CB930701, 2010CB922904, NSFC (WML) under grants Nos. 10934010, 112111076, 61227902, NSFC-RGC (WML) under grants Nos. 11061160490, 1386-N-HKU748/10 and NSF CHE-0349189 (JHR).

\end{document}